\documentclass[aps,amsfonts,nofootinbib,article,twocolumn,showpacs]{revtex4}
\usepackage{amsmath}
\usepackage{amsfonts}
\usepackage{amssymb}

\def\openone{\leavevmode\hbox{\small1\kern-3.8pt\normalsize1}}
\def\N{\leavevmode\hbox{ Z \kern-8 pt\normalsize{Z}}}
\def\openone{\leavevmode\hbox{\small1\kern-3.8pt\normalsize1}}
\def\openJ{\leavevmode\hbox{J \kern-9.5pt\normalsize J}}
\def\openS{\leavevmode\hbox{ S \kern-9.3pt\normalsize S}}
\newcommand{\bb}{\begin{equation}}
\newcommand{\ee}{\end{equation}}
\newcommand{\eqb}{\begin{eqnarray}}
\newcommand{\eqf}{\end{eqnarray}}

\usepackage{color}

\begin{document}

\title{Origin of conical dispersion relations}

\author{Sergio A. Hojman}
\email{sergio.hojman@uai.cl} \affiliation{Departamento de Ciencias,
Facultad de Artes Liberales, Facultad de Ingenier\'{\i}a y Ciencias,
Universidad Adolfo Ib\'a\~nez, Santiago, Chile,\\ and Departamento
de F\'{\i}sica, Facultad de Ciencias, Universidad de Chile,
Santiago, Chile,\\ and Centro de Recursos Educativos Avanzados,
CREA, Santiago, Chile.}

\begin{abstract}
A mechanism that produces conical dispersion relations is presented.
A Kronig Penney one dimensional array with two different strengths
delta function potentials gives rise to both the gap closure and the
dispersion relation observed in graphene and other materials. The
Schr\"{o}dinger eigenvalue problem is locally invariant under the
infinite dimensional Virasoro algebra near conical dispersion points
in reciprocal space, thus suggesting a possible relation to string
theory.
\end{abstract}

\pacs{03.65.-w}

\maketitle
\section{Introduction}

Dirac massless fermions dispersion relations have been observed in
graphene and other materials. In spite of the extensive literature
published on the subject, see for instance
\cite{ref1,ref2,ref3,ref4,ref5} and references therein, there does
not seem to exist a cogent
understanding of the essence of this remarkable feature.\\

Materials in which this dispersion relation arises have different
structures and their symmetry groups do not match but share a common
feature: They all have more than one lattice point per unit cell.
Although this feature seems
to be necessary it does nor appear to be sufficient.\\

Here we present a mechanism that produces conical dispersion
relations. A modified Dirac-Kronig-Penney one dimensional array with
two different strengths delta function potentials gives rise to both
the gap closure and the dispersion relation observed in graphene and
other materials. The condition that the (first neighbors) structure
factor vanishes defines the position of the conical dispersion
points in reciprocal space and
produces both the gap closure and conical dispersion relations.\\

The Schr\"{o}dinger eigenvalue problem is locally invariant under
the infinite dimensional Virasoro algebra near conical dispersion
points in reciprocal space, thus suggesting a relation to string
theory.\\

\section{The Model}

Consider the usual Dirac-Kronig-Penney problem (use the notation defined in \cite{maslov})
described by the potential $V_{DKP}(x)$
\begin{equation}
V_{DKP}(x)= U \ \sum\limits_{n=-\infty}^{n=+\infty} \delta (x-na).\
\label{DKP1}
\end{equation}
The associated Schr\"{o}dinger equation is
\begin{equation}
-\frac{{\hbar}^2}{2m} {\psi}'' + V_{DKP}(x){\psi}= E{\psi}. \
\label{sch1}
\end{equation}
In this case, as it is well known, the translational symmetry
transformation is
\begin{equation}
x'=x+a \ \label{t1}
\end{equation}
The eigenvalue problem implies the following dispersion relation
\begin{equation}
\cos ka - \cos qa - \frac{m\ U \ a}{{\hbar}^2} \frac{\sin qa}{qa}=0\
, \label{dr1}
\end{equation}
where
\begin{equation}
q\equiv \frac{\sqrt{2mE}}{{\hbar}}\ . \label{q}
\end{equation}
It is convenient to define the functions $f_1(k,a)$ and
$g_1(q,a,U)$ by
\begin{equation}
f_1(k,a) \equiv \cos ka  , \label{f1}
\end{equation}
and
\begin{equation}
g_1(q,a,U) \equiv  \cos qa + \frac{m\ U \ a}{{\hbar}^2} \frac{\sin
qa}{qa}\ , \label{g1}
\end{equation}
respectively, so that the dispersion relation (\ref{dr1}) may be
equivalently written as
\begin{equation}
f_1(k,a) = g_1(q,a,U) . \label{dr2}
\end{equation}
It is very well known that (\ref{dr2}) gives rise to energy gaps due
to the fact that while $-1 \le f_1(k,a) \le +1$, there are local
minima of $g_1(q,a,U)$ which are inferior to $-1$ and local maxima
that exceed $+1$. We could therefore say that there are two kind of
energy gaps, ``maxima gaps'' and ``minima gaps''.\\

Consider now the modified potential $V_M(x)$, defined by
\begin{equation}
V_M(x)=\sum\limits_{n=-\infty}^{n=+\infty}( U \ \delta (x-2na)+V\
\delta (x-(2n+1)a)).\ \label{VM}
\end{equation}
The associated Schr\"{o}dinger equation is
\begin{equation}
-\frac{{\hbar}^2}{2m} {\psi}'' + V_M(x){\psi}= E{\psi}. \
\label{sch1}
\end{equation}

In this case, clearly, the translational symmetry transformation is
\begin{equation}
x'=x+2a \ \label{t2}
\end{equation}
The dispersion relation, in this case, turns out to be
\begin{eqnarray}
&\ &\cos 2ka - \cos 2qa - \frac{m\ U \ a}{{\hbar}^2} \frac{\sin
2qa}{qa}-\\ \nonumber &\ & -\frac{m\ V \ a}{{\hbar}^2} \frac{\sin
2qa}{qa} - \frac{2 m^2\ UV \ a^2}{{\hbar}^4} \frac{(\sin qa)^2}{q^2\
a^2}=0\, \label{dr3}
\end{eqnarray}
Introduce now the functions  $f_2(k,a)$ and $g_2(q,a,U,V)$ defined
by
\begin{eqnarray}
f_2(k,a) \equiv \cos 2ka, \label{f2}
\end{eqnarray}
and
\begin{eqnarray}
&\ &g_2(q,a,U,V) \equiv  \cos 2qa + \frac{m\ U \ a}{{\hbar}^2}
\frac{\sin 2qa}{qa}\\ \nonumber &\ & +\frac{m\ V \ a}{{\hbar}^2}
\frac{\sin 2qa}{qa} +  \frac{2 m^2\ UV \ a^2}{{\hbar}^4} \frac{(\sin
qa)^2}{q^2\ a^2}\, \label{g2}
\end{eqnarray}
respectively, so that the dispersion relation (\ref{dr3}) may be
equivalently written as
\begin{eqnarray}
f_2(k,a) = g_2(q,a,U,V). \label{dr4}
\end{eqnarray}
It is interesting to note that
\begin{eqnarray}
f_2(k,a) = 2 [f_1(k,a)]^2 -1, \label{id1}
\end{eqnarray}
while
\begin{eqnarray}
g_2(q,a,U,V) = 2 g_1(q,a,U) g_1(q,a,V) -1. \label{id2}
\end{eqnarray}

\section{The V=U case}

We will study the dispersion relation (\ref{dr4}), in general, but
it is convenient to start the analysis by considering the case $V=U$
(which is inadequate from a physical standpoint because in that case
the symmetry translation vector is given by (\ref{t1}) instead of
(\ref{t2})).\\

In this case,
\begin{eqnarray}
g_2(q,a,U,U) = 2 [g_1(q,a,U)]^2 -1. \label{id2}
\end{eqnarray}
Therefore, the dispersion relation (\ref{dr2}) implies that
\begin{eqnarray}
f_2(k,a) = g_2(q,a,U,U), \label{drUU}
\end{eqnarray}
is satisfied. The converse statement is false.\\

It is, of course, clear that both $f_2(k,a)\ge -1$ and
$g_2(q,a,U,U) \ge -1$. Therefore, $-1$ is the minimum value attained by both functions.\\

The minimum values are reached for  $f_2(k,a)$ when
\begin{eqnarray}
f_1(k_r,a)=0,
\end{eqnarray}
and for $g_2(q,a,U,U)$ when
\begin{eqnarray}
g_1(q_s,a,U)=0,
\end{eqnarray}
at points ($k_r= (2r+1)\pi/2a, q_s$). \\

Therefore, some ``minima gaps'' disappear in this scheme. The
dispersion relation at generic minima points $(k_r, q_s)$ defined by
\begin{eqnarray}
f_2(k_r,a) = g_2(q_s,a,U,U)=-1, \label{drUU-1}
\end{eqnarray}
is degenerate because both the functions {\it {and}} their derivatives coincide
at points $(k_r, q_s)$. In fact, the functions derivatives vanish at
those points, due to the fact that the equality is reached at the
minima of the functions \cite{h2}, i.e, we have that at the points $(k_r,
q_s)$
\begin{eqnarray}
\frac{d f_2(k,a)}{d k}\vert _{k=k_r} = 4\frac{d f_1(k,a)}{d k}
f_1(k,a) \vert _{k=k_r}=0 \label{ddr1}
\end{eqnarray}
and
\begin{eqnarray}
&\ & \frac{d g_2(q,a,U,U)}{d q}\vert _{q=q_s} = \\ \nonumber &\ & =
4 \frac{d g_1(q,a,U)}{d q} g_1(q,a,U)   \vert _{q=q_s} = 0,
\label{ddr2}
\end{eqnarray}
also.\\

This degeneracy gives rise to local invariance under the Virasoro algebra at the conical dispersion relations points $(k_r,
q_s)$, in reciprocal space ($k$-space, see below). This feature is discussed in Section V.\\

Therefore, the dispersion relations in the vicinity of the minima
read
\begin{eqnarray}
&\ & \frac{d^2 f_2(k,a)}{d k^2}\vert _{k=k_r}(k-k_r)^2 = \\
\nonumber &\ & = \frac{d^2 g_2(q,a,U,U)}{d q^2}\vert _{q=q_s}
(q-q_s)^2 , \label{cones}
\end{eqnarray}\\ \\
Note that both
\begin{eqnarray}
\frac{d^2 f_2(k,a)}{d k^2}\vert _{k=k_r}= 4 a^2>0
\end{eqnarray}
and
\begin{eqnarray} \frac{d^2 g_2(q,a,U,U)}{d q^2}\vert _{q=q_s} = 4 \left(\frac{d g_1(q,a,U)}{d q}\right)^2 \vert _{q=q_s} >0
\end{eqnarray}
because the points $(k_r, q_s)$ define both functions minima.\\ \\

Define
\begin{eqnarray}
g_1' \equiv \frac{d g_1(q,a,U)}{d q} \vert _{q=q_s} \ ,
\end{eqnarray}
\begin{eqnarray}
f_2'' \equiv \frac{d^2 f_2(k,a)}{d k^2}\vert _{k=k_r}(=4 a^2) \ ,
\end{eqnarray}
\begin{eqnarray}
g_2'' \equiv  \frac{d^2 g_2(q,a,U,U)}{d q^2}\vert _{q=q_s} = 4
(g_1')^2 \ ,
\end{eqnarray}
\begin{eqnarray}
\Delta k \equiv k-k_r \ ,
\end{eqnarray}
\begin{eqnarray}
\Delta E \equiv \frac{{\hbar}^2 q^2}{2m} - \frac{{\hbar}^2
q_s^2}{2m} \ ,
\end{eqnarray}
to get the conical dispersion relation
\begin{eqnarray}
\Delta E= \pm \sqrt{\frac{f_2''}{g_2''}}\frac{{\hbar}^2 q_s}{m}
\Delta k = \pm {\frac{a}{ g_1'}}\frac{{\hbar}^2 q_s}{m} \Delta k =
\pm {\hbar} v_F \Delta k, \label{cones1}
\end{eqnarray}\\

where the Fermi velocity $v_F$ is given by

\begin{eqnarray}
v_F \equiv \sqrt{\frac{f_2''}{g_2''}}\frac{{\hbar} q_s}{m}= {\frac{a}{ g_1'}}\frac{{\hbar} q_s}{m}. \label{fv}
\end{eqnarray}\\

\section{The general case}

Note that for $V \neq U$, one may write
\begin{eqnarray}
&\ &g_2(q,a,U,V) \equiv g_2(q,a,U,U) + \\ \nonumber &\ & +\
\frac{2 m(V-U)a}{{\hbar}^2} \frac{\sin qa}{qa} g_1(q,a,U),
\label{g2UV}
\end{eqnarray}\\
therefore
\begin{eqnarray}
&\ & g_2(q_s,a,U,V) = g_2(q_s
,a,U,U)= \\ \nonumber &\ & =
f_2(k_r,a) = -1 , \label{gUV=gUU}
\end{eqnarray}

It is interesting to realize that the corresponding condition
$g_1(q'_s,a,V)=0$ (exchanging $V$ and $U$) also implies
$g_2(q'_s,a,U,V)=-1$, as it
should. In what follows, we write everything in terms of the the $g_1(q_s,a,U)=0$ condition only,
understanding that similar results are reached in points $(k_r,q_s)$ and points $(k_r,q'_s)$.\\

The function  $g_2(q,a,U,V)$ does not, in general, attain a minimum
at the points $q_s$. In fact,
\begin{eqnarray}
&\ & \frac{d g_2(q,a,U,V)}{d q}\vert _{q=q_s} =
\\ \nonumber &\ & = \frac{2 m(V-U)a}{{\hbar}^2}
 \frac{\sin q_sa}{q_sa} \frac{d g_1(q,a,U)}{d q} \vert _{q=q_s}\\ \nonumber &\ & =
\frac{2 m(V-U)a^2}{{\hbar}^2 q_s a}\left(-1+\frac{\sin 2 q_s a}{ 2
q_s a}\right)\neq 0,
\end{eqnarray}
in general. We nonetheless get conical dispersion relations both for low energies $q_sa\rightarrow 0$ and for high energies
(as compared to the potential strength difference) $\frac{2 m(V-U)a^2}{{\hbar}^2 q_s a}<<1$.\\

The tight binding approach ($\frac{m\ U \ a}{{\hbar}^2}>>1$ {\it
{and}} $\frac{m\ V \ a}{{\hbar}^2}>>1$) also yields interesting
results. In fact, in that case one may study the behavior of
$g_1(q_n,a,U)=0$ (or $g_1(q'_n,a,V)=0$) near the zeroes of ($\sin qa
/ qa$) \cite{maslov}, i.e., at the points $q_n a= n \pi + (-1)^n {\delta}_n, n\in
\{\mathbb{Z}-\{0\}\}$
\begin{eqnarray}
g_1(q_n, a, U)= (-1)^n +\frac{m\ U \ a}{{\hbar}^2} \frac{{\delta}_n}{n \pi}.
\label{tb1}
\end{eqnarray}
The requirement  $g_1(q_n,a,U)=0$ implies
\begin{eqnarray}
{\delta}_n= (-1)^{n+1} \frac{n \pi{\hbar}^2}{m\ U \ a},
\label{deltaun}
\end{eqnarray}
while for $g_1(q'_n,a,V)=0$ one gets
\begin{eqnarray}
{\delta}'_n= (-1)^{n+1} \frac{n \pi{\hbar}^2}{m\ V \ a}.
\label{deltavn}
\end{eqnarray}
Note that due to the tight binding condition both ${\delta}_n<<1$
and ${\delta}'_n<<1$. Therefore, for $U \sim V$ the difference
(${\delta}_n-{\delta}'_n$) is of the order ${{\delta}_n}^2$, i.e.,
to first order,  $q_n=q'_n$, which means that $g_2(q_n, a, U, V)
=-1$, thus $g_2(q_n, a, U, V)$ attains minima at $q=q_n$ in the
tight binding approximation, giving rise to conical dispersion
relations at those points.

The energy spectrum is readily computed, to get
\begin{eqnarray}
{E_n}= \frac{n^2 {\pi}^2 {\hbar}^2}{2 m a^2}
\left(1-\frac{{\hbar}^2}{mUa} \right)^2 \label{energy1}
\end{eqnarray}
or
\begin{eqnarray}
{E_n}= \frac{n^2 {\pi}^2 {\hbar}^2}{2 m a^2}
\left(1-\frac{{\delta}_1}{\pi} \right)^2  \simeq  \frac{n^2 {\pi}^2
{\hbar}^2}{2 m a^2} \left(1-2 \frac{{\delta}_1}{\pi} \right)
\label{energy2}
\end{eqnarray}
to first order in ${\delta}_1$.

\section{Local Virasoro invariance in reciprocal space}

Define $F(k,q,a,U,V)$ by
\begin{eqnarray}
F(k,q,a,U,V)\equiv f_2(k,a)-g_2(q,a,U,V). \label{F1}
\end{eqnarray}
The function $F(k,q,a,U,V)$ and its first derivatives with respect
to $k$ and $q$ vanish at ($k_r= (2r+1)\pi/2a, q_s$) (($k_r=
(2r+1)\pi/2a, q'_s$) ) whenever the conditions for conical
dispersion relations are met (see sections III and IV). As a matter
of fact, $(k_r, q_s)$ and $(k_r, q'_s)$ define saddle points
for the function $F(k,q,a,U,V)$.\\

For conical dispersion relations $F(k_r+\Delta k,q_s+\Delta
q,a,U,V)$ may be written as
\begin{eqnarray}
&\ & F(k_r+\Delta k,q_s+\Delta q,a,U,V)\simeq \\ \nonumber &\ & \simeq  \frac{{\partial}^2
F}{{\partial} k^2} \vert _{k=k_r} {\Delta k}^2
+\frac{{\partial}^2 F}{{\partial} q^2} \vert _{q=q_s} {\Delta q}^2,
\label{F2}
\end{eqnarray}
because the function and its first derivatives vanish at those
points. Moreover, the product of the second derivatives of $F$ is negative.\\

Therefore, the eigenvalue equation $F(k,q,a,U,V)=0$ in the vicinity
of conical points ($k_r= (2r+1)\pi/2a, q_s$) reduces to
\begin{eqnarray}
\frac{{\partial}^2 F}{{\partial} k^2} \vert _{k=k_r} {\Delta k}^2
+\frac{{\partial}^2 F}{{\partial} q^2} \vert _{q=q_s} {\Delta
q}^2=0, \label{F3}
\end{eqnarray}
which defines a two dimensional ``light'' cone associated to the
Fermi velocity $v_F$. The Fermi velocity is proportional to the (square
root) of the (negative of) the ratio of the second derivatives of the function
$F(k,q,a,U,V)$  at the conical points as showed in \eqref{fv}. A conformally invariant
metric near the conical points may be defined in terms of $v_F$. Two
dimensional conformal transformations leave the dispersion relation (\ref{F3})
near conical points invariant.\\

 It is a widely known fact that such structures are
invariant under the action of the (infinite dimensional) Virasoro
algebra \cite{vir}. This symmetry gives rise to local gauge
invariance of the Schr\"{o}dinger wave function around conical
points. This fact encourages the search of a relationship of this problem
to string theory (the results will be presented in a forthcoming article \cite{h1}).\\

In three dimensional models (two space dimensions) \cite{ht1}, the symmetry is
reduced to the (ten generators) three dimensional conformal algebra.

\section{Conclusions}

I have presented a mechanism that produces conical dispersion
relations consistent with behavior observed in graphene and other
materials \cite{ref1,ref2,ref3,ref4,ref5}. The mechanism assumes the
existence of more than one lattice point per unit cell. The
condition to achieve conical dispersion relations is equivalent to
require that the nearest neighbors structure factor vanish
\cite{ht1}. The Schr\"{o}dinger eigenvalue problem near conical
points happens to be invariant under the Virasoro algebra which
hints a relationship between this problem and string theory. This
feature is currently under study \cite{h1}. In a forthcoming paper
\cite{ht1}, a bidimensional generalization which includes Hubbard
model (tight binding) calculations, the ideas presented here are
explored and extended. The invariance near conical points, in two
dimensional arrays, is reduced to the ten dimensional conformal
group in three (spacetime) dimensions.

\begin{acknowledgments}

I wish to sincerely express my gratitude to Andr\'es Concha, Paula
Mellado for bringing this problem to my attention and to Felipe
Torres for many enlighting discussions.

\end{acknowledgments}


\begin{thebibliography}{17}

\bibitem{ref1}P. R. Wallace, Phys. Rev. {\bf 71}, 622 (1947)
\bibitem{ref2} Cheol-Hwan Park, Li Yang, Young-Woo Son, Marvin L.
Cohen and Steven G. Louie, Nature Physics {\bf 4}, 213,
(2008)
\bibitem{ref3}  A.Varykhalov, D. Marchenko, J. S\'anchez-Barriga, M. R. Scholz, B.
Verberck, B. Trauzettel, T. O. Wehling, C. Carbone, and O. Rader,
Phys. Rev. X {\bf 2}, 041017 (2012)
\bibitem{ref4} Daniel Malko, Christian Neiss, and Andreas G\"{o}rling, Phys. Rev. B {\bf 86}, 045443 (2012)
\bibitem{ref5}A. H. Castro Neto, F. Guinea, N. M. R. Peres, K. S. Novoselov and A.
K. Geim, Rev. Mod. Phys, {\bf 81}, 109 (2009)
\bibitem{ht1} Sergio A. Hojman and Felipe Torres, ``Origin of Dirac Massless Fermions'' (in preparation)
\bibitem{maslov} D. L. Maslov PHYZ426: ``Dirac-Kronig-Penney Model'' (January 25,2012)
http://www.phys.ufl.edu/{\raise.17ex\hbox{$\scriptstyle\sim$}}maslov/\\
phz6426/phz6426\_dkp.pdf (currently unavailable)
\bibitem{h2} Sergio A. Hojman, ``Evaluation of derivatives of positive semidefinite functions'' (in preparation)
\bibitem{vir} M. A. Virasoro, Phys. Rev. D {\bf 1}, 2933 (1970)
\bibitem{h1} Sergio A. Hojman, ``Conical dispersion relations and string theory'' (in preparation)


\end{thebibliography}
\end{document}